\documentclass[prd,12pt,preprintnumbers,nofootinbib,showpacs]{revtex4}
\usepackage{epsfig,amsmath,amssymb}
\def\Q1{{\bf Q}_1}

\begin{document}
\preprint{\bf CERN-TH/2003-081}\preprint{\bf FIRENZE-DFF -
402/03/2003}
%\baselineskip = 1\baselineskip \setcounter{eqletter}{0}
%\setcounter{equation}{0} \setcounter{section}{0}
%\setcounter{subsection}{0} \setcounter{subsubsection}{0}
%\setcounter{figure}{0}

\title{Ladder-QCD at finite isospin chemical potential}
\author{A. Barducci, G. Pettini, L. Ravagli}

\affiliation{Department of Physics, University of Florence and
INFN
Sezione di Firenze \\
Via G. Sansone 1, I-50109, Sesto F.no, Firenze, Italy}
\email{barducci@fi.infn.it, pettini@fi.infn.it,
ravagli@fi.infn.it}

\author{R. Casalbuoni\footnote{On leave from the
Department of Physics of the University of Florence, 50019,
Florence, Italy}} \affiliation{TH-Division, CERN, CH-1211 Geneva
23, Switzerland}
 \email{casalbuoni@fi.infn.it}

\begin{abstract}                % DON'T CHANGE THIS LINE
We use an effective QCD model (ladder-QCD) to explore the phase
diagram for chiral symmetry breaking and restoration at finite
temperature with different $u,d$ quark chemical potentials. In
agreement with a recent investigation  based on the
Nambu-Jona-Lasinio model, we find that a finite pion condensate
shows up for high enough isospin chemical potential
$\mu_{I}=(\mu_{u}-\mu_{d})/2$. For small $\mu_{I}$ the phase
diagram in the $(\mu_B,T)$ plane shows two first order transition
lines and two critical ending points.
\end{abstract}
  \maketitle

%**********************************************************************
\section{INTRODUCTION}
\label{int}
%**********************************************************************
In the last few years the study of QCD at finite density has
become rather important. In particular it has been established
that at zero temperature and in the high density limit a color
superconducting phase exists (for a review see
\cite{Rajagopal:2000wf}). From a phenomenological point of view
there are two areas where finite density is relevant. The first
area is the realm of compact stellar objects where the central
density can reach values up to ten times the saturation density
$\rho$, with $\rho\approx 0.14$ fm$^{-3}$  evaluated as the
inverse of the volume of a sphere of radius $1.2$ fm. Since the
temperature of a compact star is much smaller than the typical
color superconducting gap (of order of tens of ${\rm MeV}$) one
can safely consider the limit $T\to 0$. The second area   can be
found in heavy ion physics. However, in this case, color
superconductivity is not relevant given the large entropy per
baryon produced in heavy ion collisions. But here another
important feature of QCD might be relevant. According to
different models
\cite{Barducci:1989wi,kunihiro,Halasz:1998qr,Berges:1998rc} the
phase diagram of QCD in the plane $(\mu_B,T)$ exhibits a
tricritical point. Since this point should be located at moderate
density and temperature there is some possibility of observation
in heavy ion experiments. Another important point in heavy ion
physics is the fact that in the experimental setting there is a
non zero isospin chemical potential, $\mu_I$. Studies at finite
$\mu_I$ have been the object of several papers
\cite{Bedaque:1999nu,Alford:2000ze,Buballa:1998pr,Steiner:2000bi,
Neumann:2002jm,Steiner:2002gx} but mainly in the regime of low
temperature and high baryon chemical potential. The first
complete study of the phase diagram in the three-parameter space
$(\mu_B,T,\mu_I)$ has been made in the context of a Random Matrix
model \cite{Klein:2003fy}. It has been found that the first order
transition line ending at the tricritical point of the case
$\mu_I=0$ actually splits in two first order transition lines and
correspondingly two crossover regions are present at low values
of baryon chemical potential. The existence of this splitting has
also been shown in the context of a Nambu-Jona-Lasinio  model in
\cite{Toublan:2003tt}. It should also be noticed that in
\cite{Frank:2003ve} the NJL model has been augmented by the
four-fermi instanton interaction relevant in the case of two
flavors. These authors have found that the coupling induced by
the instanton interaction between the two flavors might wash
completely the splitting of the first order transition line. This
happens for values of the ratio of the instanton coupling to the
NJL coupling of order 0.1-0.15.

In this paper we will consider the effect of a finite isospin
chemical potential in a model (ladder-QCD) where the existence of
a tricritical point was shown several years ago
\cite{Barducci:1989wi}. The reason of doing this analysis in a
model different from the NJL model is due to the fact that QCD at
finite baryon density is difficult to be studied on the lattice
(however it should be noticed that recently a new technique has
been proposed \cite{Fodor:2001au} and a first evaluation of the
tricritical point has been given in \cite {Fodor:2001pe}). It is
therefore important to study certain features in different models
in order to have a feeling about their universality. For instance
the existence of a tricritical point seems to enjoy such a
characteristic. What we are presenting here is a preliminary
study, and therefore we will restrict the analysis at small
isospin chemical potential. The interest for this topic is due to
results from lattice simulations and effective theories which
show the existence of a phase transition at finite $\mu_I$
\cite{Kogut:2002tm,Kogut:2002zg,Son:2000xc,Toublan:1999hx}. We
also ignore the effects from color superconductivity, since these
are present only for temperatures lower than some tens of ${\rm
MeV}$.

In our model all the three flavors are present, and the relevant
instanton effects would give rise to a six-fermi contact
interaction. Therefore it is not clear if these effects will wash
out the splitting as in the two flavor case \cite{Frank:2003ve}.
We will consider this problem in a future work.

%****************************************************************
\section{THE MODEL (REVISITED)}
\label{sec:physics}
%****************************************************************
In this Section we will review a model that  was used several
years ago to describe  the chiral phase of QCD both at zero
temperature \cite{Barducci:1984pr,Barducci:1987gn} and at finite
temperature and density \cite{Barducci:1989wi,Barducci:1993bh}.
This model is an approximation to QCD based on the evaluation of
the effective potential at two-loop level and on a
parametrization of the self-energy consistent with the OPE
results. The effective action that we evaluate is a slight
modification (see for instance \cite{Barducci:1989wi}) of the
Cornwall-Jackiw-Tomboulis action for composite operators
\cite{Jackiw:cv,Cornwall:vz}. We recall here the major steps of
this calculation. We start from the Cornwall-Jackiw-Tomboulis
formula

\begin{equation}
\Gamma [S]~=~-~\Gamma_2[S]~+~{\rm
Tr}\left[S\frac{\delta\Gamma_2}{\delta S}\right]~-~{\rm Tr}~{\rm
ln}\left[S_0^{-1}+\frac{\delta\Gamma_2}{\delta S}\right]+ {\rm
counterterms} \label{eq:gammacjt}\end{equation} where the free
fermion inverse propagator is

\begin{equation}
S_0^{-1}(p)~=~i{\hat p}~-~m \label{eq:freeprop}
\end{equation}
and $\Gamma_{2}$ is the sum of all 2PI diagrams with propagator
$S$, which has to coincide with the exact fermion propagator at
the absolute minimum of $\Gamma$. Thus $S$ is the dynamical
variable in this variational approach. However it turns out useful
to trade $S$ for
\begin{equation}
\Sigma~=~-\frac{\delta\Gamma_2}{\delta S} \label{eq:sdeq}
\end{equation}
which  coincides with the fermion self-energy at the minimum of
$\Gamma$.

\begin{center}
\begin{figure}[htbp]
\includegraphics[width=3cm]{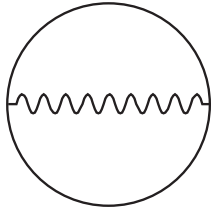}
\caption{\it The two-loop diagram needed to evaluate $\Gamma_2$.}
\label{fig:gam2}
\end{figure}
\end{center}

In the present model (ladder-QCD) we will make the very rough
approximation of evaluating $\Gamma_2$ at the lowest order. That
is, we evaluate $\Gamma_2$ at two-loops with one gluon exchange.
The relevant Feynman diagram is given in Fig. \ref{fig:gam2}. It
turns out that this approximation works rather well
phenomenologically (see for instance \cite{Barducci:1987gn}).
Therefore, at this order the effective action is simply
\cite{Barducci:1987gn}

\begin{eqnarray}
\Gamma[\Sigma]~&=&~-{\rm Tr}~{\rm ln}\left(S_0^{-1}-\Sigma\right)
-\frac 1 2~{\rm Tr}\left(S~\Sigma\right)+ {\rm c.t.}\nonumber\\
&=&~-{\rm Tr}~{\rm
ln}\left(S_0^{-1}-\Sigma\right)~+~\Gamma_{2}\left[\Sigma\right]+{\rm
c.t.} \label{eq:simplesp}\end{eqnarray}
As in refs. \cite{Barducci:1984pr,Barducci:1987gn}, we use the
following parametrization for $S$

\begin{equation}
S(p)~=~iA(p){\hat p}~+~B(p)~+~i\gamma_{5} C(p) \label{eq:parprop}
\end{equation}
Then, by working in the Landau gauge, it is possible to show that
no renormalization of the wave function is required and also that
the  Ward identity at this order  is satisfied by taking the free
quark-gluon vertex and the free gluon propagator. We also
consider the so-called rigid case, where the strong coupling $g$
is considered fixed at $p^2=M^2$, where $M$ is a convenient  mass
scale to be fixed later. In this way the relation between the
scalar and pseudoscalar contribution to $\Sigma$ and the terms
$B,~C$ in Eq. (\ref{eq:parprop}) can be easily inverted, leading
to the following expression for the effective action, fully
expressed in terms of $\Sigma$ (\cite{Barducci:1987gn})
\begin{equation}
\Gamma[\Sigma]=\Gamma_{2}[\Sigma]+\Gamma_{\rm log}[\Sigma]
\end{equation}
where, separating $\Sigma=\Sigma_s+i\gamma_5\Sigma_p$

\begin{equation}
\Gamma_2[\Sigma]=-\frac{8\Omega_4 N_{c}\pi^2} {3g^2C_2}\int
\frac{d^4q} {(2\pi)^4}{\rm
tr}~[\Sigma_s(q^2)\square~\Sigma_s(q^2)+
\Sigma_p(q^2)\square~\Sigma_p(q^2)]\label{eq:7}
\end{equation}
with $\Omega_{4}$ being the four-volume and $C_{2}=4/3$ for
$N_{c}=3$  the quadratic Casimir of $SU(3)_c$. Besides $M$, also
$g$ is a  parameter of the model. The one-loop term is
\begin{equation}
\Gamma_{\rm log}[\Sigma]=-{\rm Tr}~{\rm
log}\left(S_0^{-1}-\Sigma\right)= -{\rm Tr}~{\rm log}~\left[i{\hat
p}- m -\Sigma_s(p^2)-i\gamma_5\Sigma_p(p^2)\right]
\label{eq:unloop}
\end{equation}
where the scalar and pseudoscalar parts of the dynamical variable
are matrices in $SU(3)$ flavor space (as well as $m$), related to
the scalar and pseudoscalar quark condensates through the
following equation

\begin{equation}
\Sigma_s(p^2)+i\gamma_5\Sigma_p(p^2) =({s}+i\gamma_5 {p})~f(p^2)
\label{eq:ope}
\end{equation}
The function $f(p^2)$ which contains the momentum dependence of
the self-energy will be discussed in a moment. The fields

\begin{equation}
\langle s_{ab}\rangle=-\frac{3C_2g^2} {4NM^3}\langle
\bar{\Psi}_{a}\Psi_b\rangle \label{eq:10}
\end{equation}
\begin{equation}\label{condpseud}
\langle p_{ab}\rangle=-\frac{3C_2g^2} {4NM^3}\langle \bar{\Psi}_a
i\gamma_5\Psi_b\rangle
\end{equation}
will be determined by minimizing the effective action.  In Eq.
(\ref{eq:unloop}), $m$ is the mass matrix in flavor space which is
taken diagonal
$$
 m=\left(
\begin{array}{ccc}
m_u & 0 & 0 \\
0  & m_d & 0\\
0 & 0 & m_s
\end{array}
\right)
$$
The function $f(p^2)$ will be chosen requiring that it goes to a
constant for $p\to 0$ and as $1/p^2$ (mod log terms)  for large
values of $p$ as suggested by the OPE expansion. By introducing a
dimensionless variable $x^2=p^2/M^2$ we will consider the
following family of functions

\begin{equation}
\frac{f_N(x^2)}{M}= \displaystyle{\frac{1+x^2+x^4+...+x^{2N-2}}
{1+x^2+x^4+...+x^{2N-2}+x^{2N}}}
\label{eq:smoothnew}\end{equation} In the limit $N\to\infty$ we
get the function used in \cite{Barducci:1984pr,Barducci:1987gn}
\begin{equation}
{f(x^2)\over
M}=\theta\left(1-x^2\right)~+~\theta\left(x^2-1\right)~
\displaystyle{{1\over x^2}} \label{eq:stepold}\end{equation}
Notice that for $x\to 0$ we get
\begin{equation}
{f_{N}(x^2)\over M}\sim 1 - x^{2N}+.. \label{eq:expandf}
\end{equation}
Now, let us consider for simplicity the chiral limit and zero
chemical potentials. In this case it is simple to get the
mass-shell condition from the one loop term in
Eq.~(\ref{eq:unloop}) (see for instance \cite{Barducci:1989wi})
\begin{equation}
p^2+\langle s\rangle ^2 f^2_N=0 \label{eq:mashell}
\end{equation}
where $\langle s\rangle$ is the field proportional to the scalar
condensate (see Eq.~(\ref{eq:10})). If we want to recover, at
least in the infrared regime, a free particle-like dispersion
relation (as for instance happens in four fermion theories), we
see from Eqs.~(\ref{eq:expandf}) and (\ref{eq:mashell}) that we
need $N\geq 2$. In this paper we will choose $N=2$. Notice that
the choice $N=1$ would lead to the following dispersion relation
in the limit of small momenta
\begin{equation}
p^2(1-2\langle s\rangle ^2)+M^2\langle s\rangle ^2+...=0
\end{equation}
This might give rise to problems  in the broken phase where the
coefficient of $p^2$ could become negative. However no
difficulties arise for the determination of the critical points
where $\langle s\rangle\simeq 0$. On the contrary the equation of
state  could be affected.

We can thus evaluate explicitly the effective potential
\begin{equation}
V={\Gamma\over\Omega_{4}}
\end{equation}
which is UV-finite in the chiral limit, whereas it needs to be
properly renormalized in the massive case . We have employed the
following normalization condition
\begin{equation}\label{eq:condnorm}
\frac{\partial V}{\partial (m_a \langle
\bar{\Psi}_a\Psi_a\rangle)}\Big|_{min}=1
\end{equation}
In the chiral limit, this requirement is equivalent to the
Adler-Dashen relation (see for instance \cite{Barducci:1987gn}).
Here we will require the validity of this equation at the values
of the quark current masses.\\
By defining $\alpha_{a}=m_{a}/M$ and
\begin{equation}{\chi}_{a}=-
\frac{g^2}{3M^3}\langle\bar\Psi_a\Psi_a\rangle \label{eq:hatchi}
\end{equation}
the normalization condition, using
Eq. (\ref{eq:10}), can be written as
\begin{equation}
\frac{\partial V_a}{\partial (\alpha_a
\chi_a)}\Big|_{min}=-{3M^{4}\over 2\pi^{2}} c = -{3 M^{4}\over
g^2(M)}\label{eq:condorm1}
\end{equation}
where we have introduced the parameter
\begin{equation}c\equiv\frac{2\pi^2}{g^2(M)}\end{equation}
The chemical potential and the temperature dependence are
introduced  following standard methods \cite{Dolan:qd} (see for
example \cite{Barducci:1989wi}). In particular the chemical
potential is introduced from the very beginning  via the usual
substitution $p^{\nu}\rightarrow (p_0+i\mu,{\vec{p}})$ in  ${\hat
{p}}$ appearing in the Dirac operator in Eq.~ (\ref{eq:unloop}).
On the other hand the temperature dependence is introduced by
substituting to $p_0$ the Matsubara frequency $\omega_n=(2n+1)\pi
T$ in all the $p_0$ dependent terms appearing in the effective
action. The reason for this asymmetrical treatment is that the
$p$ dependence in the self-energies becomes relevant only at
$p\gg M$ whereas, as we shall see we will be interested in
chemical potentials lower than $M$.

\section{Results at finite temperature and density}

In order to get the effective action we need to calculate the
determinant of the operator appearing in Eq. (\ref{eq:unloop}). We
set to zero the strange quark chemical potential $\mu_{s}=0$ and
define $\mu_{I}=(\mu_{u}-\mu_{d})/2$ and
$\mu_B=(\mu_{u}+\mu_{d})/2$. The operator is given by the
following  $3\otimes 3$ matrix in flavor space:

\begin{equation}
\begin{pmatrix}
i(\omega_{n}+i\mu_{u})\gamma_{0}+i{\vec p}\cdot{\vec\gamma} -
F_{u}  &~~~~ \rho f_2\gamma_{5}& 0 \cr  -\rho f_2\gamma_{5}&
i(\omega_{n}+i\mu_{d})\gamma_{0}+i{\vec p}\cdot{\vec\gamma} -
F_{d} &0 \cr 0 & 0 &i\omega_{n}\gamma_{0}+i{\vec
p}\cdot{\vec\gamma} - F_{s} \label{determinant}
\end{pmatrix}
\end{equation}
where we have defined $F_{a}=m_{a}+ f_{2}\chi_{a}$; $~a=u,d,s$ and
with $f_{2}$ given in Eq. (\ref{eq:smoothnew}) with $N=2$. Also
$p^2=\omega_n^2+|\vec p\,|^2$. Here $\rho$ is related to the
charged pion condensate
\begin{equation}\rho=-\frac{g^2}{6M^3}
(\langle \bar u\gamma_5 d\rangle-\langle\bar d\gamma_5 u\rangle)
\end{equation}

We have directly set to zero the hyper-charged condensates in the
strange sector, since we do not expect the formation of such
condensates for $\mu_{s}=0$. The strange sector thus factorizes
and the determination of the phase diagram for approximate chiral
symmetry restoration is performed by studying the behavior of the
light $\langle {\bar u}u\rangle$ and $\langle {\bar d}d\rangle$
quark condensates, independently on the strange quark condensate.
To evaluate the effective action, we perform the sum over the
Matsubara frequencies which solve the mass-shell condition given
by the vanishing of the determinant in Eq.~(\ref{determinant})
using standard methods \cite{Dolan:qd}. Although formally
straightforward, the calculation is a very hard numerical task.
Actually, at each integration step on $|{\vec p}|$, we have to
solve a twentieth order algebraic equation in $\omega_{n}$ in the
$u-d$ sector. The part relative to the strange quark is obviously
easier to deal with. The coefficients of this equation  depend
also on the parameters
$\chi_u,\chi_d,\rho,\mu_u,\mu_d,m_{u},m_{d}$.

Before discussing the results at finite density and temperature
let us review how we fix the parameters appearing in the effective
action. This is done by looking at $T=\mu=0$. The parameters that
we have to fit are $c$, $M$, $m_u$, $m_d$, $m_s$. These are
obtained by using as input parameters the following physical
quantities $m_{\pi^\pm}$, $m_{K^\pm}$, $m_{K^0}$, $f_\pi$ and
$f_K$.  The results of the fit are given in Table I, whereas the
values of the input experimental quantities, together with the
result we get from the fit procedure are given in Table II.
\begin{table}[htbp]
\begin{center}
$$
\begin{array}{|c|c|}
\hline
{\rm {\bf Parameters}} & {\rm {\bf Fitted ~ Values}}\\
\hline
M & 529 ~ {\rm MeV}\\
\hline
c & 1.0\\
\hline
m_u & 4.4 ~{\rm MeV}\\
\hline
m_d & 6.2 ~{\rm MeV}\\
\hline
m_s & 110 ~{\rm MeV}\\
\hline
\end{array}
$$
\end{center}
\caption{{\it Fit of the parameters.}} \label{Fitpar}
\end{table}

\begin{table}[htbp]
\begin{center}
$$
\begin{array}{|c|c|c|}
\hline
{\rm{\bf Input~parameters}}&{\rm{\bf Fitted~Values}}&{\rm{\bf Experimental~Values}}\\
\hline
m_{\pi^{\pm}} & 139~{\rm MeV} & 139.6~{\rm MeV}\\
\hline
m_{K^{\pm}} &494~{\rm MeV} & 493.7~{\rm MeV}\\
\hline
m_{K^{0}} &499~{\rm MeV} & 497.7~{\rm MeV}\\
\hline
f_{\pi} & 92 ~{\rm MeV} & 92.4~{\rm MeV}\\
\hline
f_{K} & 105 ~{\rm MeV} & 113~{\rm MeV}\\
\hline
\end{array}
$$
\end{center}
\caption{{\it Comparison between the values of the input
parameters as obtained from the fit  and the experimental
results.}} \label{Valexp}
\end{table}

With these values of the parameters, we find that at $T=\mu=0$ the
quark condensate in chiral limit has the value
\begin{equation}\langle{\bar\psi}\psi\rangle_{0}=-(248~{\rm MeV})^3\end{equation}
 whereas in
the massive case \begin{equation}\langle{\bar
u}u\rangle=-(251~{\rm MeV})^3,~~~\langle{\bar
d}d\rangle=-(253~{\rm MeV})^3,~~~\langle{\bar
s}s\rangle=-(305~{\rm MeV})^3\end{equation}  and, by defining the
constituent quark masses a la Politzer \cite{Pol}
\begin{equation}
M_{const}=\bar{\Sigma}(p^2=4M_{const}^2)
\end{equation}
we get
\begin{equation}\label{Masseff2}
M_s=385~{\rm MeV},\ \ \ \ \ \ \  M_{u,d}=256\ \rm{MeV}
\end{equation}
where, here and in the following the light quarks have been
taken degenerate with mass ${\hat m}=(m_{u}+m_{d})/2=5.3~{\rm MeV}$.\\
From the general study at $T=\mu=0$ and in the chiral limit, one
finds also that in order to break the chiral symmetry one must
have
\begin{equation}
c<1.37
\end{equation}
This condition is satisfied by our choice of parameters.

\begin{figure}[htbp]
\begin{center}
\includegraphics[width=14cm]{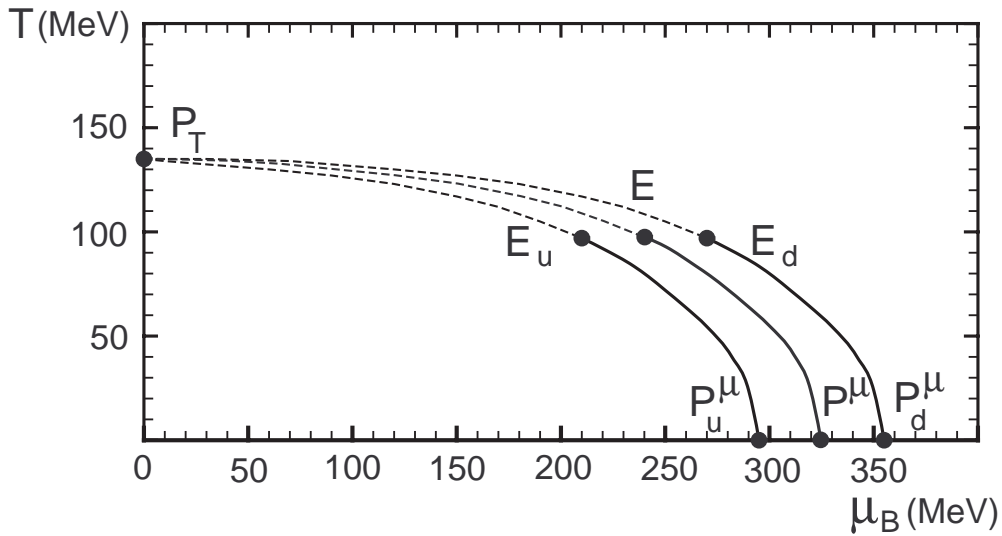}
\end{center}
\caption{\it Phase diagram for chiral symmetry in the
$(\mu_{B},T)$ plane. For $\mu_{I}=0$ (central line), the
cross-over transition line starts from the point $P_{T}=(0,135)$
and ends at the point $E=(240,97)$. The line between $E$ and the
point $P^{\mu}=(325,0)$ is the line for the first order transition
with discontinuities in the $\langle{\bar u}u\rangle$ and
$\langle{\bar d}d\rangle$ condensates. For $\mu_{I}=30~{\rm MeV}$
(side lines), the two cross-over transition lines start from the
point $P_{T}=(0,135)$ and end at the points $E_{u}=(210,97)$ and
$E_{d}=(270,97)$. The lines between $E_{u}$ and the point
$P^{\mu}_{u}=(295,0)$ and  between $E_{d}$ and the point
$P^{\mu}_{d}=(355,0)$ are the lines for the first order
transitions with discontinuities in the $\langle{\bar u}u\rangle$
and $\langle{\bar d}d\rangle$ condensates respectively.}
\label{fig:diafaison2}
\end{figure}

When $\mu_I=0$ it was shown that in the model discussed in
\cite{Barducci:1989wi} there is a tricritical point in the chiral
limit. The model we are presenting here is essentially the same
model with some slight modifications, as for instance, the choice
of the function $f_2(x)$. However the tricritical point is still
present as it can be seen from Fig. \ref{fig:diafaison2} (central
line), obtained with the choice of parameters of Table I.

A complete analysis of the full three parameter space
$(\mu_B,T,\mu_I)$ has not yet been completed especially in
relation with the pion (and hyper-charged) condensate. We have
examined the case $T=\mu_B=0$ and we have found that there is a
phase transition indicated by a finite pion condensate starting
at $\mu_I=70$ MeV. Since in our model $m_\pi\approx 140$ MeV we
agree with the results found in the literature
\cite{Kogut:2002tm,Kogut:2002zg,Son:2000xc,Toublan:1999hx,Klein:2003fy}.
We found also that for $m=0$ the critical $\mu_I$ is zero.
Therefore we will limit our considerations at small isospin
chemical potential, say $\mu_I=30$ MeV, where we expect the pion
condensate $\rho$ to vanish. In this situation the determinant in
Eq.~ (\ref{determinant}) factorizes and the effective action is
given by a sum of three independent terms, one for each  flavor.
Therefore the action is the same as for $\mu_I=0$, with each
flavor evaluated at its own chemical potential. It follows that
the light flavor terms  show the same tricritical structure
exhibited from the central line in Fig. \ref{fig:diafaison2} for
$\mu_I=0$. Consequently the phase diagram we obtain for a small,
fixed isospin chemical potential ($\mu_{I}=30~{\rm MeV}$) is
described by the two side lines in Fig. \ref{fig:diafaison2} with
each flavor $u$ and $d$ showing the same structure as the central
line but with a split chemical potential
$\mu_{u,d}=\mu_B\pm\mu_I$. Notice also that although the Figure
extends up to zero temperature, this part should not be taken too
seriously due to the existence of color superconductivity.

%*******************************************************************
\section{CONCLUSIONS}
\label{sec:conclu} In this paper we have discussed an approximate
model of QCD (ladder-QCD) at finite temperature and densities. In
particular we have considered the experimentally important
situation of a non-vanishing isospin chemical potential. This
situation has been already explored by various authors previously
and we confirm, in particular, the results found in
\cite{Klein:2003fy} and \cite{Toublan:2003tt} about the splitting
of the first order transition line in the plane $(\mu_B,T)$, for
small $\mu_I$. As pointed out in \cite{Toublan:2003tt} this
result could be relevant for ion physics since the first order
transition line is split symmetrically with respect to the
original line at $\mu_I=0$. This implies a reduction of the value
of the baryon chemical potential at the tricritical point of an
amount given by $\mu_I$, making easier the possibility of
discovering it experimentally. Since it is very difficult to
perform first principle analysis of QCD at finite baryon density,
we think that it is important to show that certain features as
the existence of the tricritical point and the possible splitting
of the first order transition line are common to several models.
This suggests that these features might possess some universal
character. However we should stress that the result of the
splitting of the first order transition line is strictly related
to the factorization in flavor space. For instance, in the two
flavor case, the four-fermi interaction due to the instanton
effects leads to a mixing of the flavors that, if sufficiently
large, might wash out the mixing. We think that this point needs
further analysis in the more complete scheme with three flavors.

%***********************************************************************
%\begin{acknowledgments}
%***********************************************************************
%We are grateful to
%\end{acknowledgments}

%***********************************************************************

%***************

\end{document}